\documentstyle[aps,multicol,epsfig]{revtex}
\title{Simple Model for Attraction between Like-Charged  Polyions}
\author{Jeferson J. Arenzon}
\address{Instituto de F{\'\i}sica, Universidade Federal
do Rio Grande do Sul\\ Caixa Postal 15051, 91501-970, Porto Alegre, RS,
Brazil  \\ E-mail: arenzon@if.ufrgs.br}    
\author{J\H{u}rgen F. Stilck}
\address{Instituto de F{\'\i}sica, Universidade Federal
Fluminense  \\ Av. Litor{\^a}nea, s/n$^o$, 24210-340, Niter{\'o}i, RJ,
Brazil \\
E-mail: jstilck@if.uff.br}       
\author{Yan Levin\footnote{Corresponding author}}
\address{Instituto de F{\'\i}sica, Universidade Federal
do Rio Grande do Sul\\ Caixa Postal 15051, 91501-970, Porto Alegre, RS,
Brazil \\ E-mail: levin@if.ufrgs.br}

\date{January 8, 1999}
\begin{document}

%\pacs{
%  \Pacs{05}{70.Ce}{Thermodynamic functions and equations of state}
%    \Pacs{61}{20.Qg}{Structure of associated liquids: electrolytes, molten
%    salts,etc}
%      \Pacs{61}{25.Hq}{Macromolecular and polymer solutions; polymer melts;
%      swelling }
%        }       
\maketitle
\begin{abstract}
We  present a simple model for the possible mechanism of appearance of
attraction between like charged  polyions inside a polyelectrolyte solution.
The attraction is found to be short ranged, and exists only in presence of
multivalent counterions.  The attraction is produced by the correlations in the 
condensed layers of counterions surrounding each  polyion,   
and appears only if the number of condensed  counterions exceeds the threshold,  
$ n > Z/2 \alpha $, where $\alpha$ is the valence of counterions and $Z$ is
the polyion charge. 
 
\end{abstract}
%\pacs{
%  \Pacs{05}{70.Ce}{Thermodynamic functions and equations of state}
%  \Pacs{61}{20.Qg}{Structure of associated liquids: electrolytes, molten
%salts,etc}
%  \Pacs{61}{25.Hq}{Macromolecular and polymer solutions; polymer melts;
%swelling } 
%  } 

%\pacs{PACS: 05.70.Ce; 61.20.Qg; 61.25.Hq}

%\bigskip
%PACS.05.70.Ce - Thermodynamic functions and equations of state
%PACS.61.20.Qg - Structure of associated liquids: electrolytes, molten salts, etc.
%PACS.61.25.Hq - Macromolecular and polymer solutions; polymer melts; swelling
%\bigskip
%\bigskip

\begin{multicols}{2}   
\narrowtext

Polyelectrolyte solutions and charged colloidal suspensions present an
outstanding  challenge to  modern statistical mechanics. One of the reasons for
the great difficulty 
in achieving an understanding of these complex systems 
is due to the intricate  role  played by the counterions (microions) which  
counterbalance
the much bigger charge of the polyions (macroions).  Since the number density of 
counterions  is  so much greater than that of polyions  it is the
counterions that dominate the thermodynamic properties of 
polyelectrolyte solutions at low densities.  
The heterogeneity combined with
the long-range Coulomb force  makes polyelectrolytes 
almost impossible to study by the traditional methods of 
liquid state theory.   It is,  however, exactly this complexity that is responsible
for the  richness of the behaviors encountered in polyelectrolyte
solutions and  charged suspensions.  

One of the most fascinating results of the subtle interplay of various  
interactions is the appearance of attraction between two
like-charged macromolecules inside a solution or a suspension.   
This attraction is purely electrostatic and is not
a result  of some additional short-range van der Waals  force.  
The attraction has  been observed  in simulations of 
strongly asymmetric electrolytes \cite{pat80},
as well as  a number of experiments \cite{blo91}.  
The exact  mechanism responsible for
this unusual phenomena is still not understood, although some theories
attempting  to  explain its basis, have been recently advanced \cite{HaLiu,oos68}.  
At this time, we believe, there exists a great need to explore various  simple 
models which might
help shed additional light on the origin of this attraction.  

Our discussion will be restricted to rodlike polyelectrolytes,
a good example of which is an aqueous solution of DNA segments.
Let us consider the simplest model of such a polyelectrolyte solution.
The polyions shall be represented by rigid cylinders of net charge $-Zq$
distributed uniformly, with
separation  $b$,  along the major axis of a macromolecule.  The counterions
shall be  treated as small rigid  spheres of charge  $\alpha q$
located at the center.  The counterions
can be monovalent, divalent, or  trivalent, for simplicity
however, we shall restrict our solution  to contain only one of the above types
of  counterions. An appropriate number of coions is also present in the solution 
to keep an overall charge neutrality.  The solvent shall be modeled by a uniform 
medium of a dielectric
constant $D$. It has been argued by Manning that in the limit of very large $Z$
and infinite  dilution, 
a certain number of counterions will condense onto the polyions,
thus renormalizing their effective charge.  From a simple phenomenological
argument Manning determined the number of condensed counterions to be
$ n_c=(1-1/\alpha \xi) Z/  \alpha$  for  $\xi>1/\alpha$ and $n_c=0$ for  
$\xi<1/ \alpha$, where $\xi=q^2/Dk_BTb$ \cite{man69}.   Note that this result 
subtly depends on the order of
the limits to be taken,  first $Z \rightarrow \infty$ and {\it then} the 
infinite dilution.
If the limits are interchanged no condensation will appear.  Recently we have extended
the Manning theory to finite concentrations and finite polyion sizes.  In this case
it is  possible to show that the counterion  association  still  persists, however,
instead of a fixed number of condensed counterions associated with each polyion,
we now find a distribution  of clusters, each composed  
of {\bf one} polyion and $1 \leq m  \leq Z/ \alpha$
associated counterions.  The distribution of cluster sizes is well  localized, and  
in the limit of large polyion charge and infinite dilution approaches a delta 
function centered on the value
proposed by Manning, $n_c$ \cite{lev96}.

Can the condensed layer of counterions be responsible
for the observed attraction between two  like charged polyions?   
To answer this question we propose the following simple model.
Consider two parallel, rodlike polyions, with $Z$ monomers, inside the 
polyelectrolyte solution.  The separation 
between two macromolecules is $d$.  If the attraction is
produced by some sort of charge-correlation  mechanism,  we expect that it should 
be short-ranged.  We shall, therefore, restrict our attention to
distances such that $d< \xi_D$, where $\xi_D$ is 
the Debye screening length.  As was mentioned  earlier, the strong  electrostatic
attraction between the  polyions and the counterions favors the formation
of clusters  composed of one  polyion and some number of associated counterions. 
For the purpose of this exposition, we shall neglect the  polydispersity of
cluster sizes and assume that both polyions have $n<Z/\alpha$ condensed counterions.  
It is important to remember that we are 
concerned with the interaction between the two polyions {\it inside} a polyelectrolyte 
solution.  As was
stressed before, an isolated polyion can confine counterions
only if it is extremely long, while inside a solution the cluster formation can 
take place with the polyions of any size \cite{lev96}.

The associated counterions are free to move
along the length of the polyions.   We define the occupation variables 
$\sigma_{ij}$,
with $i=1, 2, ...,Z$ and $j=1,2$, in such a way that $\sigma_{ij}=1$
 if a counterion is attached at $i$'th monomer of the $j$'th
polyion and $\sigma_{ij}=0$ otherwise. Since the number of condensed counterions 
is fixed by thermodynamics \cite{lev96}, the values of occupation variables obey
the constraint $\sum_{i=1}^Z \sigma_{i1} = \sum_{i=1}^Z \sigma_{i2} = n$.  
We shall
assume that the only effect of the counterion association is a local renormalization
of the monomer charge.  The Hamiltonian for this model takes a  particularly 
simple form,
\begin{equation}
{\cal H}=\frac{1}{2D}\sum_{i,i'=1}^{Z} \sum_{j,j'=1}^{2}
\frac{q^2(1-\alpha\sigma_{ij})(1-\alpha\sigma_{i'j'})}
{r(i,j;i',j')},
\label{e1}
\end{equation}
where the sum is restricted to $(i,j) \neq (i',j')$, 
$r(i,j;i',j')=b\sqrt{|i-i'|^2 + (1-\delta_{jj'})x^2}$ is the distance between
the monomers located at $(i,j)$ and $(i',j')$, 
$\delta_{jj'}$ is the Kronecker delta, and $x=d/b$. Clearly, the above model is a 
great over-simplification of physical reality.  The molecular nature of the solvent 
is not taken into account.  The counterions are assumed to be confined to the surface 
of the polyions,  while the polyions themselves are treated as completely rigid. 
Nevertheless, we believe that the simplicity of this phenomenological model, which allows us to perform exact analytic calculations, 
compensates for its abstract nature.  Evidently, if we can understand under 
what conditions the attraction can arise between two macromolecules in this 
idealization,  in the future it might pave the way to a 
more complete, physically realistic model.
 
With this disclaimer in mind, we proceed to
rewrite the Hamiltonian as $\beta {\cal H}=\xi H$,
where $\beta =  1/k_BT$ and $\xi$ is the Manning parameter defined
earlier.  The adimensional reduced Hamiltonian is
\begin{equation}
H=\frac{1}{2}\sum_{(i,j) \neq (i',j')}
\frac{(1-\alpha\sigma_{ij})(1-\alpha\sigma_{i'j'})} {\sqrt{|i-i'|^2 +
(1-\delta_{jj'})x^2}}.  
\label{e2}
\end{equation}

Using a transformation of occupation variables defined by
$\sigma'_{ij} = 1-\sigma_{ij}$ it is easy to see that the Hamiltonian
exhibits the following symmetry
\begin{equation}
H(Z,\alpha,\{\sigma\}) = (\alpha-1)^2 H(Z,\alpha',\{\sigma'\}),
\label{e3}
\end{equation}
where $n'=Z-n$ and $\alpha'=1+1/(\alpha-1)$.

The partition function is
\begin{equation}
Q={\sum_{\{\sigma_{ij}\}}}^\prime\exp(-\beta {\cal
H})={\sum_{\{\sigma_{ij}\}}}^\prime  \exp(-\xi H),
\label{e4}
\end{equation}
where the prime indicates that the occupation numbers are subjected to the
constraints of the number of condensed counterion conservation. 
The symmetry relation (\ref{e3}) leads to
invariance of the partition function, and any thermodynamic quantity derived 
from it, $Q(Z,n,\xi,\alpha)=Q(Z,Z-n,[\alpha-1]^2 \xi,1+1/[\alpha-1])$.
This property is quite useful when performing the calculations, 
since it reduces the ranges of parameters needed to study. 

It is convenient to
rewrite the partition function in terms of the variables
$y_i=\exp(-\xi/i), i=1,2,...,Z-1$,
associated with the {\it intrapolyion} interactions and
$z_i=\exp(-\xi/\sqrt{x^2+i^2}), i=0,1,...,Z-1$,
related to the {\it interpolyion} interactions. 
The partition function, for given values of $Z$ and $n$, may be
expressed as
\begin{equation}
Q=\sum_i^{N_c} \prod_{j=1}^{Z-1} y_j^{u_{ij}} \prod_{k=0}^{Z-1} 
z_k^{v_{ik}},
\label{e5}
\end{equation}
where the sum runs over all the allowed configurations of counterions. The
exponents $u_{ij}$ and $v_{ik}$ are quadratic polynomials in $\alpha$
with integer coefficients.  For not too big values of $Z$, it
is possible to generate the whole set of integers in the expression
above, thus obtaining the partition function exactly.  The force
between the polyions may now be calculated through
\begin{equation}
F=\frac{1}{b \beta}\frac{\partial \ln Q}{\partial x}.
\label{e6}
\end{equation}

%%%%%%%%%%%%%%%% figure %%%%%%%%%%%%%%%%%%%%%
\begin{figure}[h]
\centerline{\epsfig{file=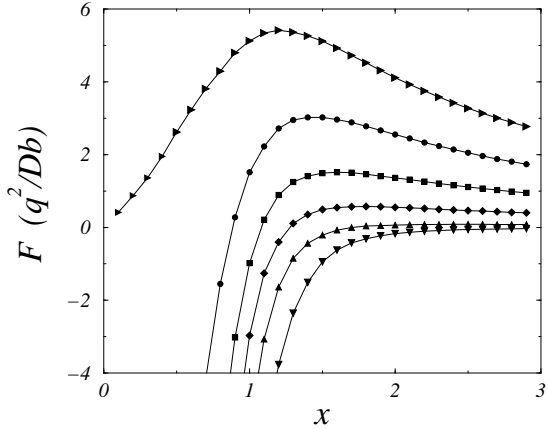,width=6cm,angle=270} }
%\vspace{1cm}
\caption{Force versus distance between polyions for $Z=20$, $\alpha=2$,
$\xi=2.283$ (corresponding to polymethacrylate) and $n=5,\ldots,10$ (from top
to bottom) in the Monte Carlo simulation. The lines are only guides to the eye.}
\label{figure1}
\end{figure}
%%%%%%%%%%%%% end of figure %%%%%%%%%%%%%%%%% 

\begin{figure}[h]
\centerline{\epsfig{file=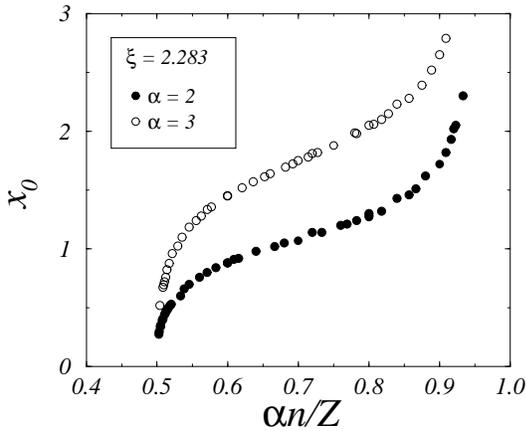,width=7cm} }
%\vspace{1cm}
\caption{Distance below which the polyion interaction becomes attractive
versus the number of condensed counterion in the Monte Carlo simulation. 
Averages are over 100 sets  and $Z$ ranges from 20 to 200.}
\label{figure2}
\end{figure}

We find that at short distances and for $\alpha > 1$ the force between the
polyions may become attractive (negative). The exact results are in full agreement 
with
the Monte Carlo simulations which can also be extended to much larger values
of $Z$, Fig. \ref{figure1}. For
$\alpha = 1$ the force is always repulsive, which is
in full agreement with the experimental evidence on absence of
attraction if only monovalent counterions are present \cite{blo91}. When the
polyion charge is completely neutralized, $n=Z/\alpha$, the force becomes 
purely attractive, as can be seen in Figs. \ref{figure1} and \ref{figure2}.
Furthermore, the exact solution and the
 simulations indicate that a critical number of  condensed counterions
is necessary for attraction to appear, see Fig. \ref{figure2}.   

%%%%%%%%%%%%%%%% figure %%%%%%%%%%%%%%%%%%%%%
%%%%%%%%%%%%% end of figure %%%%%%%%%%%%%%%%%                        

%%%%%%%%%%%%%%%% figure %%%%%%%%%%%%%%%%%%%%%
\begin{figure}[h]
\centerline{\epsfig{file=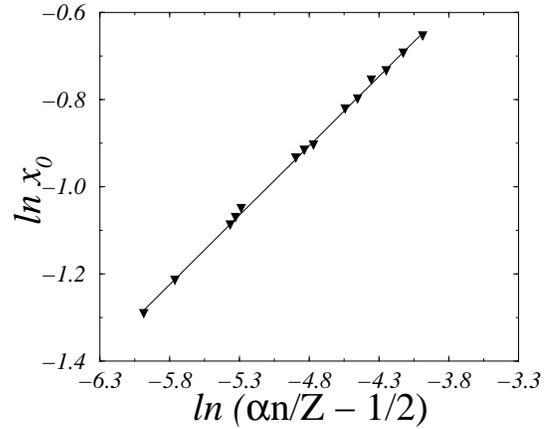,width=7cm} }
%\vspace{1cm}
\caption{Log-log plot near $n=Z/2\alpha$ for the $\alpha=2$ data in fig.
\ref{figure2}. The fit exponent is 0.32.}
\label{figure3}
\end{figure}
%%%%%%%%%%%%% end of figure %%%%%%%%%%%%%%%%%                          

To simulate the model we use a standard Monte Carlo with
particle-hole exchange, not
restricted to nearest-neighbor pairs. This non local diffusion comes from the
fact that the counterions may associate/dissociate at any point along the
polyion. After thermalizing, both the energy and the force are
measured, results being time
and sample averaged. In Fig.\ref{figure2} it is clear that the relevant
parameter is the fraction of neutralized charge, $\alpha n/Z$, the curves
for several values of $Z$ (here from 20 to 200) collapsing on an universal
function, deviations occurring only for very small $Z$.

As expected, the attractive interaction is
short-ranged and at larger distances 
the force becomes repulsive. To further explore
this point we consider the limit of small $x$.
In this case the variable $z_0$ vanishes, while for finite $\xi$
and $Z$ all the other variables $y_j$ and $z_j$ remain positive definite. If we
define $v=\min_i(v_{i,0})$, it is possible to rewrite the partition function as
$Q=z_0^v W(1+P)$, with
\begin{equation}
W=\sum_{i=1}^l\prod_{j=1}^{Z-1}y_j^{u_{ij}}\prod_{k=1}^{Z-1}
z_k^{v_{ik}},
\label{e7}
\end{equation}
and
\begin{equation}
P=\frac{1}{W}\left( \sum_{i=l+1}^{N_c}z_0^{v_{i0}-v}\prod_{j=1}^{Z-1}y_j^{u_{ij}}
\prod_{k=1}^{Z-1}z_k^{v_{ik}}\right),
\label{e8}
\end{equation}
where we suppose $v_{i0}=v$ for the first $l$ of the $N_c$
configurations. As $x \rightarrow 0$, the function $P$ vanishes, and
 we may use the approximation $Q \approx Wz_0^v$. We then notice
that $v$ corresponds to the configurations which maximize the number
of favorable horizontal interpolyion interactions,
%\begin{mathletters}
%\label{e9}
\begin{equation}
v=Z-2n\alpha %\mbox{, if $2n \leq Z$,}
\label{e9}
\end{equation}
for $\alpha\geq 2$.
%\begin{equation}
%v=(Z-2n) \alpha^2+2n\alpha-Z \mbox{, otherwise.}
%\end{equation}
%\end{mathletters}
Using the above approximation for the partition function we obtain
\begin{equation}
F \approx \frac{1}{b \beta}\left(\frac{1}{W}\frac{\partial
W}{\partial x} +\frac{\xi v}{x^2} \right).
\label{e10}
\end{equation}
It is not difficult to see that the derivative of $W$ vanishes as $x
\rightarrow 0$, so that for small $x$ we can Taylor expand the first term of
Eq. \ref{e10}.    To leading order we find
\begin{equation}
f \equiv b \beta F \approx \frac{\xi v}{x^2} - hx,
\label{e11}
\end{equation}
where $h<0$ (as $v$ vanishes) is a function of the remaining parameters of the model. 
We notice that as $x \rightarrow 0$ the force vanishes if $v=0$,
is large attractive if $v$ is negative,
and is large repulsive if $v$ is positive. In particular, in the limit
$v \rightarrow 0^-$  the equilibrium distance is
\begin{equation}
x_0 \approx \left(\frac{\xi v}{h}\right)^{1/3} \sim
\left(\frac{2 n\alpha }{Z} - 1\right)^{1/3}  .
\label{e12}
\end{equation}
This scaling behavior is  also  observed in Monte Carlo simulations, Fig. 
\ref{figure2} and \ref{figure3}.

%%%%%%%%%%%%%%%% figure %%%%%%%%%%%%%%%%%%%%%
\begin{figure}[h]
\centerline{\epsfig{file=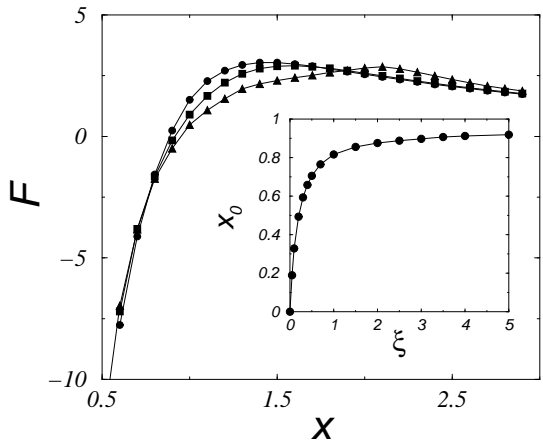,width=7cm} }
%\vspace{1cm}
\caption{Temperature dependence of the force between polyions for
$Z=20$, $n=6$, $\alpha=2$ and several values of $\xi$: 2.28 (circles),
corresponding to polymethacrylate, 4.17 (squares), corresponding
to DNA and 20 (triangles) for very low temperatures. The inset
shows the distance $x_0$ at which the attraction first appears as
a function of $\xi$. We see that for divalent counterions, $\xi=1$
serves as a division point for the low and high temperatures
regimes. Clearly for $\xi>1$ the attraction is driven by the zero
temperature mechanism!}
\label{xcsi}
\end{figure}
%%%%%%%%%%%%% end of figure %%%%%%%%%%%%%%%%%                            

We would like to stress that for a fixed number of condensed counterions
the attraction is insensitive to the temperature (see figure \ref{xcsi}).
In fact the force between two lines at zero temperature $(\xi=\infty)$
is almost exactly the same as
at finite temperature, as long as the average electrostatic
energy of interaction between two condensed counterions is greater than
the thermal energy,  $\alpha^2 q^2/D d > k_B T$, where $d$ is the average
separation between the condensed counterions.
The mechanism for attraction between two like-charged rods
is now clear \cite{rouzina}. At zero temperature the counterions on the two
rods will take on staggered configuration, i.e. if the site of
the first rod is occupied by a counterion the parallel site of the
second rod will stay vacant.  If the number of condensed counterions
is above the threshold $ n = Z/2 \alpha $ the favorable counterion-monomer
interactions will outnumber the unfavorable monomer-monomer interactions,
thus producing a net attraction at short distances. Using the threshold
value of $n$ we find that the attraction is dominated by the zero
temperature correlations as long as  
 $\xi>2/\alpha$. On the other hand applying the Manning condensation
criterion to the threshold number we see that 
the attraction
is possible {\it only} if $\xi > 2/\alpha$. We
thus come to an important conclusion: If the attraction exists
it is produced by  the zero temperature correlations.  

In view of the current interest in the possible mechanisms 
responsible for the attraction between like charged objects it is worthwhile 
to make some further comments.  A most common approach used to
study this difficult problem
relies on the field theoretic methodology similar to the one developed
in Quantum Field Theory to study Casimir forces. The partition function
is mapped onto an effective field theory which is then studied using a
loop expansion.  Due to the underlying difficulty of this process
the expansion is usually terminated at the first loop level which is
equivalent to the so called Gaussian approximation. The attraction,
in this approach, arises as a result of the correlations in the Gaussian
fluctuations\cite{HaLiu}.  Clearly from the above discussion this is
not the mechanism responsible for the attraction between the two charged
rods which was found by us.  The problem with the field theoretic approaches
used up to now is that the ground state energy was not properly treated.

Instead of allowing for a staggered configuration, which we found
to be responsible for the attraction observed, the field theoretic approaches
neglect the discrete nature of charges and uniformly
 smear
the condensed counterions over the polyion.  Clearly at zero temperature
this can only result in a repulsion!  The attraction, then, appears only as
a finite temperature correction, produced
by the  correlations in the Gaussian fluctuations of the
 counterion charge densities on the two polyions, 
the effect which is
much weaker than the one  found by us to
be responsible for the attraction. In view of our results 
 we must conclude that
 the currently used field theoretic approaches \cite{HaLiu} must be
 reexamined.

We have presented a simple model for the possible mechanism of appearance of
attraction between the like charged  polyions inside a polyelectrolyte solution.
The attraction is found to be short ranged, and is possible only in a presence of
multivalent counterions.  The attraction is produced by the correlations in 
the condensed layers of counterions surrounding each  polyion and extends uniformly from zero temperature. The attraction   
 appears only if the number of condensed  counterions exceeds the threshold,  
$n > Z/2\alpha$.  Using the counterion condensation
theory to estimate the number of associated counterions, we see that the attraction
is possible only for  polyelectrolytes with $\xi > 2/\alpha$.  This is
the fundamental 
result which has, evidently, gone unnoticed in the previous studies of this 
interesting phenomena.

%\stars  
We are grateful to Prof. J.A.C. Gallas for kindly
providing time on his Alpha station. This work was partially supported by the
Brazilian 
agencies CNPq and CAPES.

\end{multicols}     

\end{document}